\documentclass[a4paper,fleqn]{cas-sc}

\usepackage[authoryear]{natbib}
\usepackage{ulem}
\usepackage{color}

\def\tsc#1{\csdef{#1}{\textsc{\lowercase{#1}}\xspace}}
\tsc{WGM}
\tsc{QE}

\begin{document}
\let\WriteBookmarks\relax
\def\floatpagepagefraction{1}
\def\textpagefraction{.001}

\shorttitle{Showers with both northern and southern solutions}
\shortauthors{L. Neslu\v{s}an et al.}  


\title[mode = title]{Showers with both northern and southern solutions}



\author[1]{L. Neslu\v{s}an}[orcid=0000-0001-9758-1144]

\address[inst1]{Astronomical Institute, Slovak Academy of Sciences,
 05960 Tatransk\'{a} Lomnica, Slovakia}
\ead{ne@ta3.sk}



\author[2]{T. J. Jopek}

\address[inst2]{Astronomical Observatory Institute, Faculty of Physics,
         A. M. University, Pozna\'{n}, Poland}



\author[3]{R. Rudawska}

\address[inst3]{RHEA Group / ESA ESTEC, Noordwijk, The Netherlands}

\author[4]{M. Hajdukov\'{a}}

\address[inst4]{Astronomical Institute, Slovak Academy of Sciences,
                Bratislava, Slovakia}

\author[5]{G. Kokhirova}

\address[inst5]{Institute of Astrophysics, National Academy of Sciences
                of Tajikistan, Dushanbe, Republic of Tajikistan}



\begin{abstract}
Meteoroids of a low-inclination stream hit the Earth arriving from a
direction near the ecliptic. The radiant area of stream like this is
often divided into two parts: one is situated northward and the other
southward of the ecliptic. In other words, two showers are caused
by such a stream. Well-known examples of such showers are the Northern
Taurids, \#17, and Southern Taurids, \#2, or the Northern
$\delta$-Aquariids, \#26, and Southern $\delta$-Aquariids, \#5. While
the meteoroids of the northern shower collide with the Earth in the
descending node, those of the southern shower collide with our planet
in the ascending node of their orbits. Because of this circumstance and
tradition, the northern and southern showers must be distinguished.
Unfortunately, this is not always the case with meteor showers listed in
the IAU Meteor Data Center (MDC). For the same shower, some authors
reported a set of its mean parameters corresponding to the northern
shower and other authors to the southern shower. We found eleven such
cases in the MDC. In this paper, we propose corrections of these
mis-identifications.
\end{abstract}



%
\maketitle
%
\section{Introduction}
\label{sect1}

The Meteor Data Center (MDC) of the International Astronomical
Union (IAU) maintains a database of observed meteor showers
\citep{2011msss.conf....7J, 2014me13.conf..353J, 2017PSS..143....3J,
2021JIMO...49..163R}. In the database, meteor showers are sorted into
four lists according to their status. The {\it List of all showers}
contains all the showers reported to the MDC except for those which have
met conditions for their removal. Some showers in this list are regarded
as certainly existing and are classified as the ``established showers''.
The group of the established showers is provided in the {\it List of
Established Showers}, which is a sub-set of the {\it List of all
showers}. Showers that have not been given established status are placed
on the {\it Working list of showers}. In addition to these lists, a
{\it List of removed shower's data} is available in the MDC. It lists
showers or their individual representations (solutions), existence of
which is questioned and a recommendation for their removal is published
in the literature, or those which have not met, after their submission,
the MDC requirements \citep{2020P&SS..18204821J, 2022AA...01..1H}. The
IAU MDC lists of showers can be found on the web pages of the
MDC\footnote{\url{https://www.ta3.sk/IAUC22DB/MDC2022/}}. In this work,
we refer to the List of all showers.

The MDC shower database was created in AD $2007$ as a compilation of the
data reported during several decades by many authors. Before AD $2020$,
the meteor data provided were not routinely verified for their internal
consistency $-$ the correspondence of the geocentric parameters with the
corresponding elements of the orbits. Signals of problems of this nature
with data in the MDC have occasionally appeared in the literature, see
e.g. \citet{2016JIMO...44..151K}. Similarly, no tools were available to
assess whether we were dealing with a new shower or another set of
parameters of an already known shower.

Therefore, the MDC shower part database suffered from some
short\-comings. Finally, at the MDC, the decision was taken to revise
the content of the database, formulate rules and construct tools to
facilitate the control of the database and its maintenance. The first
steps have already been taken, see \cite{2022AA...01..1H}. In this work,
we continue the process of improving the MDC database, namely, we intend
to eliminate incorrect assignments of several solutions to the Northern
or Southern branches of meteor showers.

In Section~\ref{historia}, we briefly describe the historical context of
our research, in Section~\ref{teoria}, its theoretical justification.
In Section~\ref{poprawki}, we give a list of streams containing both
Northern and Southern type solutions, and suggest how to fix the
problem. Since an incorrect assignment may also occur at an
identification of meteors of an ecliptical shower, we discuss a way to
avoid this mistake in Section~\ref{discus}. The last section contains a
summary.

\section{Northern and Southern branches of meteor showers}
\label{historia}

In the interplanetary space, there can be a meteoroid stream, which
collides with the Earth in such a way that the radiant area is located
near the ecliptic. Some meteoroids of such a stream can collide with our
planet in the descending and the other meteoroids of the stream in the
ascending node of their orbits. The radiants of the meteors caused by
the descending (ascending) meteoroids are, then, located at the northern
(southern) ecliptic hemisphere of sky. We speak about the northern
(southern) shower. Although there is only a single stream in the
interplanetary space, we distinguish two showers. In practice, the
ecliptic latitude of a radiant, $\beta$, is positive for the
northern showers and negative for the southern showers.

Here, we speak about a stream colliding with the Earth in a short,
continuous arc of its orbital corridor in a specific interval of time
during the year. The Northern and Southern Taurids or Northern and
Southern $\delta$-Aquariids are examples of such
streams. In the context of this work, we do not consider the stream
colliding with the Earth in two well-separated arcs of its corridor, in
two time intervals, as the stream of, e.g., comet 1P/Halley. This stream
hits our planet as the $\eta$-Aquariids, \#31, in May (at
ascending node) and the Orionids, \#8, in October (at
descending node). We neither deal with a mechanism leading to an
occurrence of quadruple or octuple showers originating from the same
parent body. In, e.g., quadruple case, there are two pre-perihelion
intersections in the ascending and descending nodes and two
post-perihelion intersections in the ascending and descending nodes.
It means, two showers of the quadruple system are northern and two are
southern.

One can ask why is the division of the showers to northern and southern
showers necessary? Each shower is characterized by a set of mean
parameters. The set can contain the geocentric parameters as the radiant
coordinates, geocentric velocity, and solar longitude.
For some purposes, the mean orbit is also
worth to be well-known. It can be used in a search for the shower's
parent body, investigation of a resonant action, or can be useful in a
study of meteoroid physical properties, which might be influenced by an
intensive solar radiation if the perihelion distance of the stream is
small.

Considering the standard definition of orbital elements, one would face
the problem of a meaningful determination of the argument of perihelion,
$\omega$, and the longitude of ascending node, $\Omega$, in the case of
a non-divided stream, with the northern and southern particles.
The definition of these two elements is based on the ascending node.
While the meteors having the ecliptic latitude of their radiant
$\beta > 0^{\circ}$ collide with the Earth in their descending node,
those with $\beta < 0^{\circ}$ collide with our planet in the ascending
node of their orbit. Thus, both $\omega$ and $\Omega$ of the meteors of
the first group are shifted about $180^{\circ}$ in respect to the
$\omega$ and $\Omega$ of the meteors of the second group. Consequently,
a unique mean orbit of a stream containing the meteors as with
$\beta > 0^{\circ}$ as those with $\beta < 0^{\circ}$ cannot be
determined. We have to divide the stream into two groups and, at its
collision with the Earth, we must speak about two corresponding meteor
showers, northern and southern.

The above-mentioned reason for the dividing of showers to northern and
southern was obviously noticed by the meteor astronomers in an early
epoch of meteor science, although it is hard to find out when this
happened for the first time. After the World War II, \citet{Lovell1954}
however clearly distinguished between the ``northern'' and ``southern''
Taurids (the adjectives ``northern'' and ``southern'' were still
written with the lower-case initial letters,
i.e. as common adjectives; not as a part of name). He also gave the
different mean orbits for both showers, with shifted $\omega$ and
$\Omega$ (Table~111 in his book). In Fig.~147 in the book, Lovell
illustrated daily motion of the northern-Taurids mean radiant with
a curve above the ecliptic (and orbital plane of Jupiter which was also
indicated in this figure) and daily motion of mean radiant of the
southern-Taurids mean radiant with a curve below the ecliptic and
Jupiter's orbital plane. Beside the northern and southern Taurids,
Lovell also wrote about the ``northern'' and ``southern'' Arietids. We
should remember that there were not known, yet, many ecliptical showers
with both northern and southern branches in Lovell's time.

Now, we can speak about the purpose of this paper. Many showers in the
IAU MDC list were studied and their mean characteristics determined by
two or more author groups. Hereafter, we refer to a set of mean
parameters determined by one author team as to ``solution''. It is
obvious that any shower can be characterized either by the northern or
the southern solutions. No multi-solution northern shower can contain a
southern solution and no multi-solution southern shower can contain a
northern solution. In addition to the existing tradition, this
requirement was in line with the MDC's shower naming rules until August
2022 \citep[see][]{2023NewAR..9601671J}. Unfortunately, this obvious
demand is not obeyed for few showers in the MDC list. In
Section~\ref{poprawki}, we specify these showers and propose a way to
correct the defects.


\section{On the depletion of meteor radiants around the ecliptic}
\label{teoria}

A meteoroid stream is usually a continuous entity, therefore the radiant
area of a corresponding meteor shower is a continuous area. If the area
is situated near the ecliptic, i.e. it
is intersected by this plane, then a continuity could also be expected.
However, this is not true in reality, where the radiants in the part of
the radiant area close to the ecliptic are seen to be strongly depleted.
This is observed in ecliptical streams, in which, in the advanced stage
of their evolution, a characteristic split into two branches takes place
\citep{1968IAUS...33..391K}.

\begin{figure}
\centering
\includegraphics[width=0.5\textwidth]{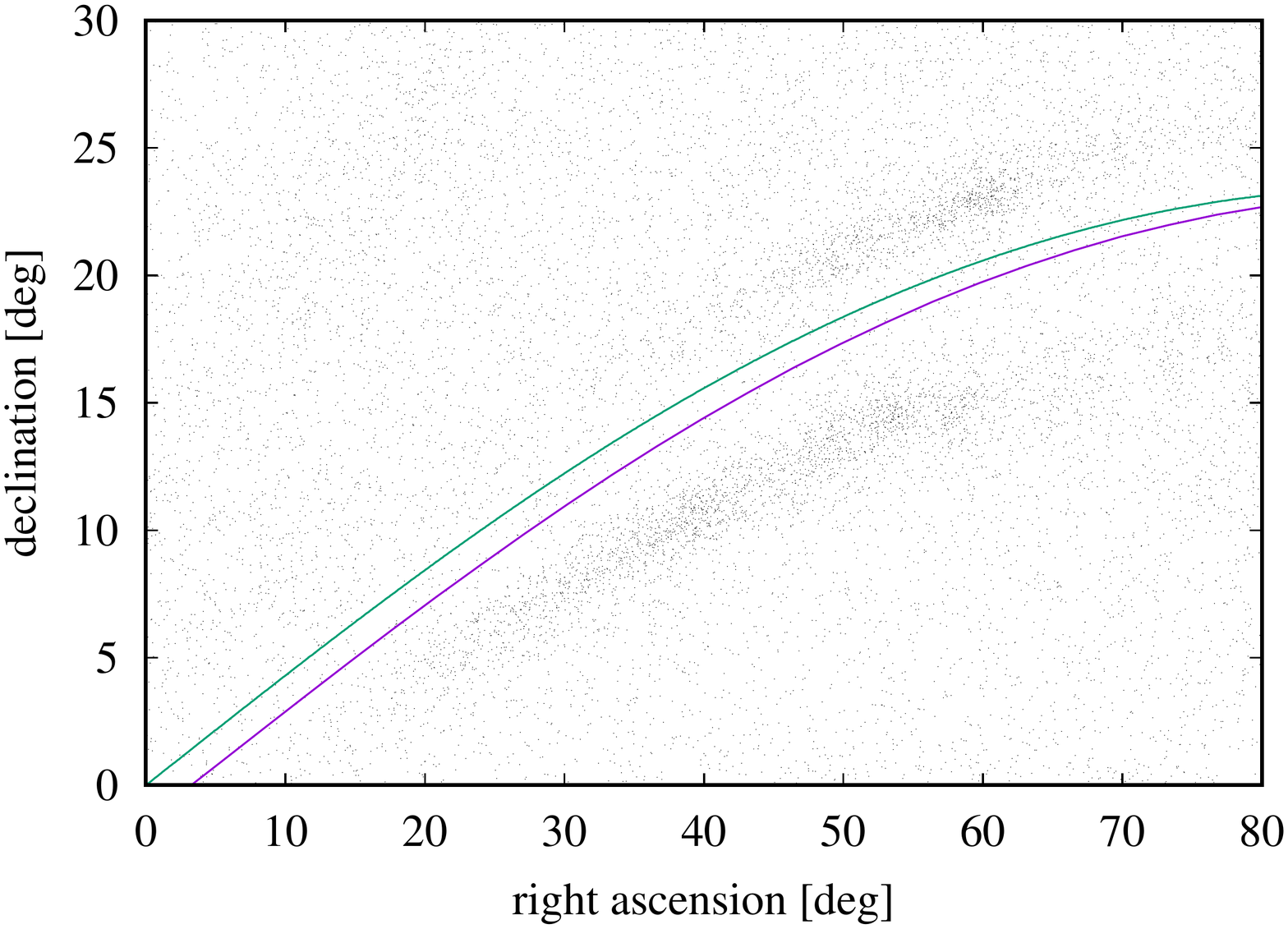}
\caption{The radiant positions of all meteors in the CAMS video
database, version 3, years 2010$-$2012, in the region of sky where the
radiants of the Northern and Southern Taurids, \#2 and \#17, are
situated. The green curve shows the behavior of the ecliptic and the
violet (lower) curve is the intersection of the orbital plane of
Jupiter with the celestial sphere in the shown region of sky.}
\label{FIGdepletion}
\end{figure}

An example of such a stream-split and depletion of orbits close to the
plane of the ecliptic can be seen in Fig.~\ref{FIGdepletion}. The figure
shows a depletion of radiants in the region of sky between the radiant
areas of the Northern and Southern Taurids, \#2 and \#17. In the figure,
the radiants of all meteors from the Cameras for All sky Meteor
Surveillance (CAMS), version 3 \citep{Gural2011, Jenniskens_etal2011,
Jenniskens_etal2016a, Jenniskens_etal2016, Jenniskens_Nenon2016,
Jenniskens_etal2016c}, records in 2010$-$2012 are
shown. We used only the part of the database, covering years
2010$-$2012, because of a better transparency of the figure. A gap
between the radiant areas of both showers caused by a single meteoroid
stream is well seen.

So, the break of the continuity of the radiant area is a real
phenomenon. It is a consequence of perturbing action on the meteoroids,
which is well-known in celestial mechanics. We recall the mechanism.
The strength of a perturbation of meteoroid orbit, due to the planets,
can be given with the help of the Lagrange equations. In the context of
meteor radiants, the perturbations of inclination, $i$, argument of
perihelion, $\omega$, and longitude of ascending node, $\Omega$, are
critical. The Lagrange equations for these three orbital elements are
\citep{Brouwer_Clemence1961}
\begin{equation}\label{LEi}
\frac{di}{dt} = \frac{\cos{i}}{na^{2}(1 - e^{2})^{1/2}\sin{i}}
 \frac{\partial \mathcal{R}}{\partial t} -
 \frac{1}{na^{2}(1 - e^{2})^{1/2}\sin i} \frac{\partial \mathcal{R}}
 {\partial \Omega},
\end{equation}
\begin{equation}\label{LEsom}
\frac{d\omega}{dt} = \frac{(1 - e^{2})^{1/2}}{na^{2}e}
 \frac{\partial \mathcal{R}}{\partial e} - \frac{\cos{i}}
 {na^{2}(1 - e^{2})^{1/2}\sin{i}} \frac{\partial \mathcal{R}}
 {\partial t},
\end{equation}
\begin{equation}\label{LEgom}
\frac{d\Omega}{dt} = \frac{1}{na^{2}(1 - e^{2})^{1/2}\sin{i}}
 \frac{\partial \mathcal{R}}{\partial t},
\end{equation}
where $\mathcal{R}$ is the disturbing function, $n$ is the mean motion,
$a$ is the semi-major axis, $e$ is the eccentricity, and $t$ is time.

In Eqs.~(\ref{LEi})$-$(\ref{LEgom}), the function $\sin{i}$ is in
the denominator of at least one fraction on the right-hand side.
Therefore, when $i \rightarrow 0^{\circ}$ or $i \rightarrow 180^{\circ}$,
then $\sin{i} \rightarrow 0$ and $|di/dt| \rightarrow \infty$,
$|d\omega/dt| \rightarrow \infty$, $|d\Omega/dt| \rightarrow \infty$.
In practice, it means that the change of $i$, $\omega$, and $\Omega$
due to a perturbation is extreme, when $i \rightarrow 0^{\circ}$ or
$i \rightarrow 180^{\circ}$. A perturbation quickly moves a meteoroid
from an orbit with a very small absolute value of inclination to an
orbit with a higher $|i|$, which is more stable. In reality, the
perturbation can result in the above-mentioned quadruple and octuple
showers.

In the Solar System, the main perturber of the meteoroids is Jupiter.
Because of this fact, the orbital elements figuring in
Eqs.(\ref{LEi})$-$(\ref{LEgom}) are implicitly assumed to be referred
to the orbital plane of this planet. Actually, the gap between the
radiant areas of the Southern and Northern Taurids, seen in
Fig.~\ref{FIGdepletion}, is symmetric in respect to the orbital plane
of Jupiter (violet curve), not to the ecliptic (green curve). This
symmetry relative to the plane of Jupiter's orbit has already been
noticed by \cite{1963SCoA....7..293T}, who, probably, for the first
time used the terms northern and southern branches for the
$\delta$-Aquariid meteoroid stream.

So, a radiant area of a stream situated at the ecliptic is
divided, by the above-described mechanism, to two parts in contrast to
a radiant area of stream situated far from the ecliptic. The latter
remains compact. The phenomenon of ``split'' radiant area of ecliptical
stream also supports the division of the ecliptical showers to the
northern and southern branches.

\section{Mixed solutions in the MDC showers and their correction}
\label{poprawki}

The discussion concerning meteor showers with mixed branches must, for
obvious reasons, concern showers for which at least two solutions are
given in the MDC. At the same time, there is assumed that all discussed
solutions concern different branches of the same stream. Whether this is
valid will be confirmed once the duplicates and false-duplicates are
revealed. Thus, our corrections may also be applied to solutions that
are false-positive in nature, i.e. that have arisen as a result of
improper grouping of different showers.

The list of the showers with the mixed, northern and southern branches
(solutions), is given in Table~\ref{tab1}. 
To correct this list, we compromise between two following rules: (1) all
the necessary corrections must be made, and (2) the minimum changes
should be made. We note that there is a need to correct only the current
list (11 showers in Table~\ref{tab1}). In the future, a mistake of this
kind should no longer appear since the mean ecliptic latitude $\beta$
will always be calculated when entering new data into the MDC database.
As a result the classification whether the solution is northern or
southern will be easy.

To make all the necessary corrections, we have to give some new names to
the showers or change already used names. This necessity is a
consequence of situation in the time when the first version of the
MDC list of showers was created. Since a central register of showers was
absent, the researchers did not know about all showers that had been
discovered, named, and referred in many papers during many decades of
meteor research. Sometimes, the same name was given to two showers.

In a creation of the MDC list, a lot of solutions were taken from
already existing lists, especially from the list in the book
of \citet{2006mspc.book.....J}. Some names in these lists were
changed in respect to the names given to the showers by the original
authors. The changed names then also occurred in the MDC list.

In some cases, a shower appeared to be similar to another shower,
therefore it was regarded as an additional
solution of the latter. The original name of the shower was, thus,
abandoned and could be used by other authors to name another new shower.
This is likely the reason why one can find two different, autonomous
showers named by the same name in the literature.

In the following, a way how to correct each of 11 problematic shower is
proposed.\\

{\bf (1) Shower \#17, the Northern Taurids} is characterized by $11$
solutions. Hereafter, we denote a particular solution of given shower by
its number (No.) assigned to the solution in Table~\ref{tab1}. Solution
No. 6 of the Northern Taurids, i.e. \#17/6 (IAUNo/No), reported by
\citet{2010Icar..207...66B} should be moved to the Southern Taurids,
\#2, since its ecliptic latitude $\beta < 0^{\circ}$ and the orbital
elements correspond to the Southern branch of Taurids. In the paper by
\citet{2010Icar..207...66B} the value of mean $\beta$ is not given,
while the mean orbital elements were computed using the radiant and
velocity observed at the time of the shower maximum. In
\citet{2010Icar..207...66B}, solution No. 6 of shower \#17 was
identified by using the 3D wavelet transform using the Sun-centered
ecliptical coordinates of radiant, solar longitude, and geocentric
velocity as an input. The Sun-centered coordinates used, only partly
compensate for the diurnal motion of the radiant, therefore, due to
the very small inclination of the average orbit ($i=0.4$), the
classification of the identified shower as Northern Taurids turned out
to be inaccurate; it should be re-classified as another solution of the
Southern Taurids, \#2.\\

{\bf (2) The Corvids, \#63}, were reported first time by
\citet{Hoffmeister1948} as a southern shower. Later,
\citet{Jenniskens_etal2016a} published another, but northern solution,
under the same shower name. 

A current, detailed inspection of Hoffmeister's solution
revealed that this solution is not regular in respect to the MDC
requirements, since only the mean radiant was determined. To give a mean
orbit, the author assumed that the Corvid stream originated
from a Jupiter-family comet and assumed the mean semi-major axis
of the stream (i) $a = 2.5\,$au and (ii) $a = 3.0\,$au. Based on this
assumption, the corresponding orbit was determined. In the book
of \citet{Hoffmeister1948}, the geocentric velocity of the Corvids
was not given. In the MDC list, the value $v_{g} = 9.1\,$km$\,$s$^{-1}$
of this parameter can be found. It was, probably, taken from the list
provided by \citet{2006mspc.book.....J}, where this value is given in
the line below Hoffmeister's solution, and belongs to another
solution of this shower, reported by \citet{Kresak_Porubcan1970}. These
authors, similarly to Hoffmeister, assumed the value of semi-major axis
$a = 3.0\,$au. Therefore, neither their solution holds the MDC criteria;
it was not selected to be included in the MDC when creating the MDC list
in 2006. 

Since Hoffmeister's solution is a hypothetical solution, we
propose to remove this solution into the List of Removed Showers. Then,
the problem with the mixed solutions will disappear and the Corvids,
\#63, will remain the northern shower.\\

{\bf (3) The $\kappa$-Aquariids, \#76}, are represented, in the current
IAU MDC list, with their northern solution reported by
\citet{Lindblad1971} and southern solution reported by
\citet{Porubcan_Gavajdova1994}. The latter was suggested to be a
solution of the $\kappa$-Aquariids by
\citet{2006mspc.book.....J}. In the original paper by
\citet{Porubcan_Gavajdova1994}, the shower with the mean parameters
identical to this solution was named as the September $\iota$-Aquariids.
Hence, we propose to remove the solution \#76/0 from the
$\kappa$-Aquariids and add the September $\iota$-Aquariids, as a ``new''
shower into the MDC Working List. Based on the old naming rules, an
IAUNo should be given to this new shower, accordingly.\\

{\bf (4) The Northern $\delta$-Cancrids, \#96}, are now represented with
6 northern and 2 southern solutions. Since the name of the shower has
the adjective ``Northern'', six northern solutions will remain to be
the solution of this shower. 

We need to deal with two southern solutions, namely those published by
\citet{Nilsson1964} and \citet{2013JIMO...41...61M}. An obvious
correction would be to move these two solutions from the Northern
$\delta$-Cancrids, \#96, to the Southern $\delta$-Cancrids, \#97, which
exists in the MDC. The shower found by Nillson was not named in his
original paper; the name Northern $\delta$-Cancrids was assigned to it
in \citet{2006mspc.book.....J}.

The second southern solution \#96/5 was obtained from an overall of 900
meteors. \citet{2013JIMO...41...61M} wrote that this solution would
normally be omitted due to a large scatter in meteor shower parameters.
The authors were aware that there is the Southern $\delta$-Cancrids
branch in the MDC with a period of activity similar to that of the
Northern $\delta$-Cancrids, and which was not detected by them as
separate shower. \citet{2013JIMO...41...61M} wrote, that maybe the high
scatter was just the side effect of this second
radiant, but they decided to report this shower anyway. It appears that
the parameters of the shower \#96/5 were obtained as a mixture of
solutions for the Northern and the Southern ones.
However, in the absence of Look-Up tables for \#96/5, we will not
unravel the resulting difficulty. 

What can be done here, to improve the MDC content, is to use a formal
approach, so in view of the negative value of the ecliptic latitude of
the mean radiant, we propose to move the solutions No. 1 and 5 of shower
\#96 to the shower \#97 as its other solutions. Showers \#96 and \#97
are the northern and southern showers of the $\delta$-Cancrid stream.\\

{\bf (5) The Southern $\delta$-Leonids, \#113}, are represented, in the
MDC list, by two solutions \citep{Sekanina1976, Terentjeva1989}.
Sekanina named this shower as $\delta$-Leonids (without the adjective
``Southern''). His solution is southern and that published by Terentjeva
appears to be a northern shower. In this case, it seems that these two
solutions are not two mixed branches; there is no shift of $\omega$ and
$\Omega$ about $\sim$$180^{\circ}$. Instead, the showers reported by
Sekanina and Terentjeva are two different showers. One can notice quite
large differences in their declination of the mean radiant
($7.4^{\circ}$ vs. $17.9^{\circ}$), mean geocentric velocity
($20.9\,$km$\,$s$^{-1}$ vs. $17.42\,$km$\,$s$^{-1}$), mean perihelion
distance ($0.580\,$au vs. $0.729\,$au), longitude of ascending node
($134.46^{\circ}$ vs. $146.36^{\circ}$), or argument of perihelion
($91.3^{\circ}$ vs. $69.0^{\circ}$).

So, the solution by \citet{Terentjeva1989} should be presented as a
different (new) shower. In the original paper by Terentjeva, it is named
as the $\alpha$-Cancrids (c). Unfortunately, name $\alpha$-Cancrids, is,
in the MDC list, used for the shower \#266 represented by a single
solution by \citet{Porubcan_Gavajdova1994} with significantly
different mean parameters in comparison with the Terentjeva's solution.

\begin{figure}
\centering
\includegraphics[width=0.7\textwidth]{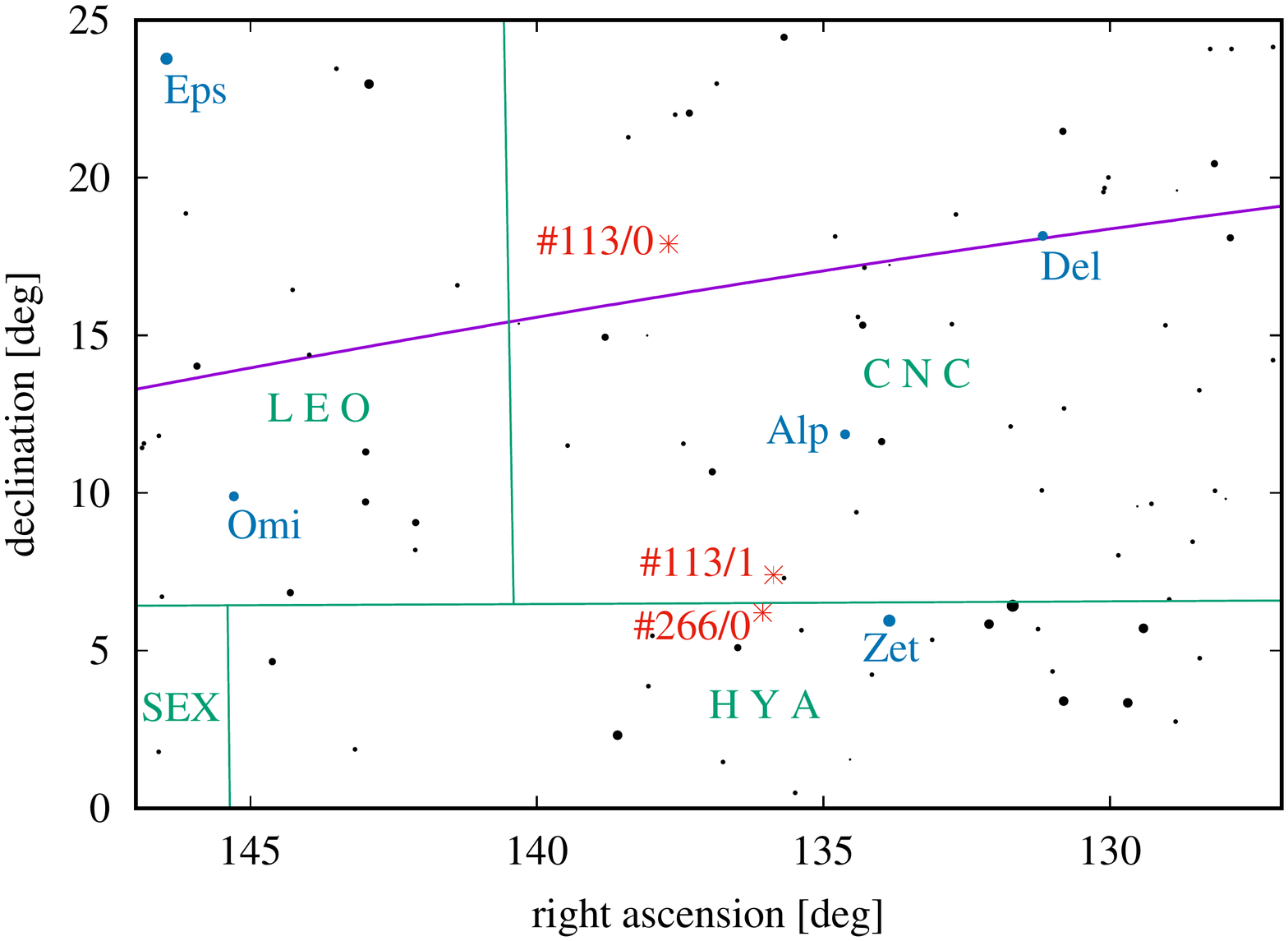}
\caption{The radiant positions (marked as red
asterisks) of the showers \#113/1 found by \citet{Sekanina1976}, \#266/0
by \citet{Porubcan_Gavajdova1994}, and \#113/0 by \citep{Terentjeva1989}.
The blue dots show some of the nearest bright stars to the
position of the mean shower radiant. The other stars are shown with
the black dots. Sekanina's radiant is located inside the
constellation Cancer, close to the star $\zeta$ Hydrae. The
radiant identified by Porub\v{c}an and Gavajdov\'{a} lies on the border
of the constellations Cancer and Hydra, also close to the $\zeta$
Hydrae. Terentjeva's shower has a radiant point well inside the
constellation Cancer clearly above the $\alpha$
Cancri star. The solid purple line indicates the position of the
ecliptic. The green lines show the borders of constellations.}
\label{fig:113-266}
\end{figure}

Interestingly, the solution for the $\alpha$-Cancrids, \#266/0,
presented by Porub\v{c}an and Gavajdov\'{a}, is similar to the solution
given by Sekanina for the Southern $\delta$-Leonids, \#113/1. This can
be seen immediately in Figure \ref{fig:113-266}. The similarity of the
positions of the radiants of showers \#133/1 and \#266/0 is very clear,
lending credence to the supposition that Sekanina made a mistake when
determining the name of its meteoroid stream. The
star $\delta$ Leonis (equatorial
coordinates being $\alpha = 168.5^{\circ}$, $\delta = 20.5^{\circ}$)
is far from the radiant located close to
$\alpha = 136^{\circ}$, $\delta = 7^{\circ}$.

Hence, in view of the above, we propose to restore the original name,
$\alpha$-Cancrids, of the solution by Terentjeva (now reported as
\#113/0 in the MDC list), but to supplement it with the adjective
Northern. A new IAUNo will be given to this ``new'' shower. Whereas,
the \#113/1 should be added as a separate solution to stream
$\alpha$-Cancrids, \#266, with the addition of Southern before the name,
since both streams have a radiant located below the ecliptic. So, if the
proposition is accepted, then shower \#266 would be renamed to the
Southern $\alpha$-Cancrids.

The shower named Southern $\delta$-Leonids will then disappear
from the MDC Working list. For the sake of archiving, its current
solutions will appear in the List of removed showers with a remark about
their moving to the other showers.\\

\begin{figure}
\centering
\includegraphics[width=0.7\textwidth]{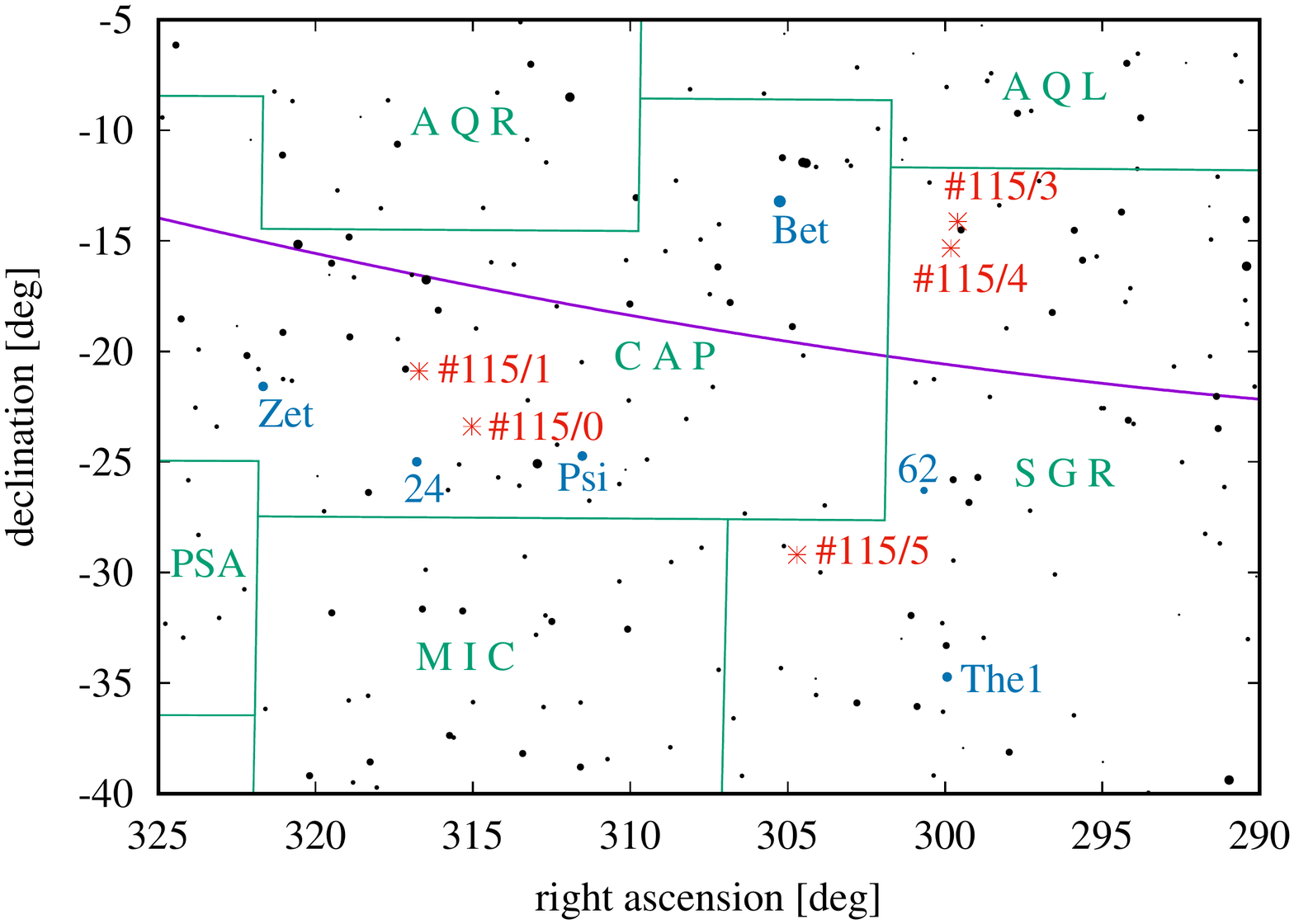}
\caption{The radiant positions (marked with red
asterisks) of five solutions of the shower \#115:
No. 3 by \citet{Sekanina1976},
No. 4 by \citet{Sekanina1973},
No. 0 by \citet{Sekanina1973},
No. 1 by \citet{Gartrell_Elford1975}, and
No. 5 by \citet{2010Icar..207...66B}.
The blue dots show some of the nearest bright stars to the
position of the mean shower radiant. The other stars are shown with
the black dots. The solid purple line indicates the position of
the ecliptic and green lines are the borders of constellations.
The figure illustrates mixed brunches of a single stream - northern
showers (solutions No. 3 and 4) and southern
showers (solutions No. 0 and 1) but also a
probable improper grouping of different showers (solution No. 5).}
\label{fig:115}
\end{figure}

{\bf (6)The Daytime Capricornids-Sagittariids (DCS), \#115}, are now
represented with two northern and three southern solutions at the MDC.
The positions of their mean radiants are shown in Fig.~\ref{fig:115}.


Three solutions of DCS in the MDC list, \#115/0, \#115/3, and \#115/4,
were reported by \citet{Sekanina1973, Sekanina1976}. Two northern
solutions, No. 3 and 4, were named by Sekanina as
Capricornids-Sagittariids (without adjective ``Daytime''). The present
name Daytime Capricornids-Sagittariids occurred in
\citet{2006mspc.book.....J}. 

Sekanina's southern solution,
\#115/0, was also added to this shower, although Sekanina named it as
the $\chi$-Capricornids. In the MDC list, the $\chi$-Capricornids,
\#420, are, however, the shower reported by \citet{Jopek_etal2010}. The
Jopek et al.'s $\chi$-Capricornid orbit is retrograde, so this shower is
fundamentally different from the Sekanina's $\chi$-Capricornids,
therefore the latter cannot become another solution of the
$\chi$-Capricornids, \#420. The second southern solution of the DCS was
published by \citet{Gartrell_Elford1975} who did not name this shower.
The third solution in the MDC list, which is also southern, was reported
by \citet{2010Icar..207...66B}.

We note that there is also a shower, which is named Daytime
$\chi$-Capricornids, \#114, in the MDC list. This shower was, again,
reported by \citet{Sekanina1973}. However, Sekanina himself named this
shower as the Equuleids, which was later changed by
\citet{2006mspc.book.....J} after he assigned the Equuleids to the
Daytime Capricornids-Sagittariids. The alternative naming of the DCS,
$\chi$-Capricornids, Daytime $\chi$-Capricornids, or Equuleids reflects
the situation in the shower names prior the MDC list was established.

We propose to replace the DCS, \#115, with two other showers: the
Southern Capricornids-Sagittariids (solutions No. 0, 1, and 5) and the
Northern Capricornids-Sagittariids (solutions No. 2 and 4). New IAUNos
will be assigned to these new showers. Name ``Daytime
Capricornids-Sagittariids'' will be deleted.\\

{\bf (7) The Daytime April Piscids, \#144}, are not a problematic case,
in fact, but we mention them because the following circumstance. In the
MDC, three solutions of the shower are given. Two clearly northern
solutions were published by \citet{2008Icar..195..317B,
2010Icar..207...66B}. The third solution is, formally, a southern
solution. It was published by \citet{K_L1967}. The latter is a curious
case. Its authors did not mention the equinox the mean angular
parameters are referred to. In the equinox 2000.0 considered in the MDC
as obligatory, the latitude of the mean radiant $\beta = -0.15^{\circ}$.
If the coordinates of the mean radiant were referred to equinox 1950.0,
then $\beta = -0.02^{o}$, however $\beta > 0^{\circ}$ in the 1900.0
equinox.

Notice that in Table~\ref{tab1} $\omega$ and $\Omega$ of the
``southern'' solution No. 0 are not shifted about $\sim$$180^{o}$ in
respect to those of the northern solutions.

Because of these circumstances we suggest to retain the solution by
\citet{K_L1967} as a peculiar northern solution of the Daytime April
Piscids, \#144. (In the MDC list, there will be inserted only a remark
about the above-mentioned circumstance.)\\

{\bf (8) The Southern May Ophiuchids, \#150}, are currently
re\-present\-ed by seven solutions. However, only southern solutions
reported by \citet{Jopek_etal2010} and \citet{Shiba2022} hold the
criteria for a regular shower. For the other solutions, no orbital
elements are given in the MDC, hence, it was not possible to verify
whether they are indeed duplicates of stream \#150.

However, solutions No. 2$-$5 were submitted in the same paper
\citep{2013JIMO...41..133M}. As the authors wrote {\it ``\#150 are an
odd case. The shower can be detected safely between May 5 and June 6
with almost 1600 shower meteors. Its rank is never lower than four, and
even though it is the strongest source in the sky for an extended period
of time, it is not possible to obtain sensible shower parameters.''}
In these circumstances, it is not clear why \citet{2013JIMO...41..133M}
decided that the published parameters represent the southern branch of
the stream. Surprisingly, but in an earlier paper, the authors correctly
classified the parameters of stream \#149/1 Northern May Ophiuchids,
\citep{Molau_Rendtel2009}.

Therefore, we think that the correct step would be to move the solutions
No. 2$-$5 of stream \#150, as additional solutions of the stream
\#149, Northern May Ophiuchids. Showers \#149 and \#150 are the northern
and southern showers of the May Ophiuchid stream.\\

{\bf (9) The Daytime $\beta$-Taurids, \#173}, are now represented by
four solutions. Of these, three are the southern solutions
\citep{Lovell1954, Cook1973,2008Icar..195..317B} and one is a
northern solution \citep{Sekanina1976}. Since the shower was reported
first time as the southern shower \citep{Lovell1954}, the three southern
solutions will be retained as the Daytime $\beta$-Taurids. The northern
solution reported by Sekanina should be regarded as a new shower. Since
Sekanina named it as the $\beta$-Taurids (without the adjective
``Daytime''), we propose to retain this name. Thus, based on the old
naming rules a new IAUNo should be assigned to this new northern shower.
This new shower and shower \#173 are, in fact, the northern and southern
daytime showers of the Taurid stream.\\

{\bf (10) The $\sigma$-Arietids, \#237}, are represented, in the MDC, by
data taken from \citet{K_L1967} and \citet{Molau_Rendtel2009}
publications. The first solution obtained by radio technique describes
the southern branch, the second one obtained by single station video
observations represents the northern branch of this shower. No orbit was
given for the northern branch, so it is not a stream whose parameters
are fully known. In \citet{K_L1967}, the authors did not provide a name
for their stream, the name $\sigma$-Arietids was proposed in
\citet{2006mspc.book.....J}. We suggest separating these two solutions
from the shower \#237 and create two showers named Northern
$\sigma$-Arietids and Southern $\sigma$-Arietids respectively. Both
solutions will obtain a new IAUNo. The stream \#237 (No. 1 and No. 2)
will be moved to the ``List of Removed Shower Data'', with an
appropriate annotation.\\

{\bf (11) The {\it h}-Virginids, \#343}, were reported first time in the
same year by \citet{SonotaCo2009, Molau_Rendtel2009}. While the SonotaCo
team solution is a southern shower, the
solution by Molau and Rendtel is northern solution. Beside the
SonotaCo-team solution, there are other six southern solutions of
$h$-Virginids in the MDC list. One of these was published by
\citet{Jenniskens_etal2016}, other four by \citet{Roggemans_etal2020},
and the last one by \citet{Shiba2022}.

It appears that the problematic northern solutions of $h$-Virginids,
\#343/1, is consistent with the solutions of other MDC shower: the
$\alpha$-Virginids, \#21. Because of this circumstance, we propose to
move solution \#343/1 of the $h$-Virginids to become one of the
solutions of the $\alpha$-Virginids. The $\alpha$-Virginids, \#21, are
probably the northern branch of the $h$-Virginids, \#343.

\begin{table*}
\caption{The showers in the IAU MDC list with both northern and southern
solutions (branches). In the columns, there are given:
``IAU'' is the IAU number of shower;
``No'' is the number of solution of given shower;
``N/S'' indicates whether the solution is northern (N) or southern (S)
branch;
``$\beta$'' are the ecliptical latitude of mean radiant (in degrees);
``$\alpha$'' and ``$\delta$'' are the equatorial coordinates of mean
radiant (in degrees);
``$v_{g}$'' is the mean geocentric velocity (in km$\,$s$^{-1}$);
``$q$'' (in au), ``$e$'', ``$\omega$'', ``$\Omega$'', and ``$i$'' are
the elements of mean orbit (the angular elements are in degrees);
``$N$'' is the number of shower meteors in given solution; and
``yr.'' is the year when the solution was published by the original author.
All angular parameters refer to J2000 reference frame.}
\begin{flushleft}
\begin{tabular}{rrcrrrrccrrrrl}
\hline \hline
IAU & No & N/S & $\beta$ & $\alpha$ & $\delta$ & $v_{g}$ & $q$ & $e$ &
 $\omega$ & $\Omega$ & $i$ & $N$ & yr. \\
\hline
\multicolumn{14}{l}{Northern Taurids, \#17}\\
  17 & 0 & N & 1.26 &  58.60 &  21.60 & 28.30 &   0.350 & 0.832 & 294.90
 & 226.20 &  3.10 &   49 & 2002 \\
  17 & 1 & N & 1.95 &  44.71 &  19.00 & 30.69 &   0.284 & 0.871 & 302.30
 & 212.78 &  2.89 &   25 & 1970 \\
  17 & 2 & N & 2.72 &  44.70 &  19.80 & 29.60 &   0.317 & 0.853 & 298.80
 & 214.10 &  3.40 &   22 & 2003 \\
  17 & 3 & N & 2.85 &  52.40 &  21.90 & 28.80 &   0.339 & 0.841 & 296.30
 & 222.70 &  3.40 &   22 & 2003 \\
  17 & 4 & N & 1.78 &  53.30 &  21.00 & 28.10 &   0.354 & 0.828 & 294.80
 & 223.80 &  2.30 &  470 & 2008 \\
  17 & 5 & N & 3.00 &  62.00 &  24.00 & 26.70 &  &  &  &  &  &  475 & 2009 \\
  17 & 7 & N & 2.52 &  48.90 &  20.70 & 28.00 &   0.355 & 0.829 & 294.60
 & 220.60 &  3.00 &  509 & 2016 \\
  17 & 9 & N & 2.32 &  49.70 &  20.70 & 29.10 &   0.321 & 0.844 & 298.60
 & 218.80 &  3.00 & 1188 & 2022 \\
  17 & 10 & N & 2.42 & 63.70 &  23.70 & 27.10 &   0.396 & 0.815 & 289.60
 & 236.20 &  2.70 & 2170 & 2022 \\
  17 & 11 & N & 2.98 & 87.40 &  26.40 & 23.80 &   0.510 & 0.767 & 276.30
 & 263.90 &  2.50 &  143 & 2022 \\
\multicolumn{14}{l}{}\\
  17 &  6 & S & $-$0.38 &  48.90 &  17.70 & 28.10 &  0.351 & 0.830 &
 115.09 &  39.10 &  0.40 & 2281 & 2010 \\
\hline
\multicolumn{14}{l}{Corvids, \#63}\\  
  63 & 1 & N & 10.16 & 205.80 &   0.20 &  8.70 &  0.999 & 0.571 & 193.70
 &  91.80 &  2.60 &   12 & 2016 \\
\multicolumn{14}{l}{}\\
  63 & 0 & S & $-$12.88 & 192.60 & $-$19.40 &  9.10 &  &  &  &  &  &  & 1948 \\
\hline
\multicolumn{14}{l}{$\kappa$-Aquariids, \#76}\\
  76 & 1 & N & 4.54 & 341.64 &  $-$2.87 & 19.00 &  0.867 & 0.705 & 229.20
 & 186.63 &  2.09 &    4 & 1971 \\
\multicolumn{14}{l}{}\\
  76 & 0 & S & $-$2.95 & 334.67 & $-$13.67 & 12.80 & 0.884 &  &  45.20
 &   0.46 &  1.01 &    3 & 1994 \\
\hline
\multicolumn{14}{l}{Northern $\delta$-Cancrids, \#96}\\
  96 & 0 & N & 1.57 & 130.00 &  20.00 & 25.67 &  &  &  &  &  &  & 1995 \\
  96 & 2 & N & 0.73 & 126.72 &  19.92 & 28.00 &   0.448 & 0.800 & 282.59
 & 298.15 &  0.30 &    7 & 1971 \\
  96 & 3 & N & 1.30 & 124.83 &  20.92 & 25.80 &   0.425 & 0.777 & 287.89
 & 293.17 &  1.20 &   27 & 1973 \\
  96 & 4 & N & 1.42 & 130.52 &  19.71 & 26.40 &   0.397 & 0.783 & 291.29
 & 297.31 &  1.50 &   37 & 1976 \\
  96 & 6 & N & 2.47 & 127.60 &  21.50 & 27.20 &   0.410 & 0.814 & 286.60
 & 290.00 &  2.70 &   74 & 2016 \\
  96 & 8 & N & 1.63 & 128.60 &  20.40 & 28.20 &   0.398 & 0.835 & 188.10
 & 296.90 &  2.30 &  243 & 2022 \\
\multicolumn{14}{l}{}\\
  96 & 1 & S & $-$2.94 & 134.19 &  14.20 & 26.70 &  0.370 & 0.770 &
 116.71 & 120.14 &  4.90 &    6 & 1964 \\
  96 & 5 & S & $-$0.40 & 131.40 &  17.60 & 27.73 &  &  &  &  &  &
 900 & 2013 \\
\hline
\multicolumn{14}{l}{Southern $\delta$-Leonids, \#113}\\
 113 & 0 & N & 1.56 & 137.70 &  17.90 & 17.42 &   0.729 & 0.664 &  69.00
 & 146.36 &  4.29 &  & 1989 \\
\multicolumn{14}{l}{}\\
 113 & 1 & S & $-$9.00 & 135.87 &  7.40 & 17.66 & 0.580 & 0.676 &  91.30
 & 134.46 &  6.40 &   37 & 1976 \\
\hline
\multicolumn{14}{l}{Daytime Capricornids-Sagittariids, \#115}\\
 115 & 3 & N & 6.40 & 299.60 & $-$14.13 & 25.10 &  0.415 & 0.758 &
  69.80 & 309.84 &  6.20 &   29 & 1976 \\
 115 & 4 & N & 5.18 & 299.81 & $-$15.33 & 29.40 &  0.314 & 0.842 &
  60.00 & 314.03 &  6.80 &   26 & 1973 \\
\multicolumn{14}{l}{}\\
 115 & 0 & S & $-$6.11 & 315.03 & $-$23.40 & 26.80 &  0.355 &  0.789 &
 242.50 & 145.07 &  6.79 &   15 & 1973 \\
 115 & 1 & S & $-$4.16 & 316.71 & $-$20.90 & 28.91 &  0.360 &  0.820 &
 246.00 & 144.66 &  4.49 &    3 & 1975 \\
 115 & 5 & S & $-$9.33 & 304.70 & $-$29.20 & 23.80 &  0.556 &  0.792 &
 270.86 & 121.00 &  7.30 &  428 & 2010 \\
 \hline
\end{tabular}
\end{flushleft}
\label{tab1}
\end{table*}

\addtocounter{table}{-1}
\begin{table*}
\caption{$-$ continuation.}
\begin{flushleft}
\begin{tabular}{rrcrrrrccrrrrl}
\hline
 IAU & No & N/S & $\beta$ & $\alpha$ & $\delta$ & $v_{g}$ & $q$ & $e$ &
 $\omega$ & $\Omega$ & $i$ & $N$ & yr. \\
\hline
\multicolumn{14}{l}{Daytime April Piscids, \#144}\\
 144 & 1 & N & 3.54 &   3.80 &   5.50 & 28.90 &   0.256 & 0.831 &  50.10
 &  24.70 &  4.70 &  397 & 2008 \\
 144 & 2 & N & 3.10 &   4.90 &   5.50 & 29.20 &   0.249 & 0.837 &  49.49
 &  26.00 &  4.50 & 2608 & 2010 \\
\multicolumn{14}{l}{}\\
 144 & 0 & S & $-$0.15 &   7.64 &   3.14 & 28.91 &  0.220 & 0.830 &
  45.00 &  30.26 &  0.51 &   34 & 1967 \\
\hline
\multicolumn{14}{l}{Southern May Ophiuchids, \#150}\\
 150 & 2 & N & 5.06 & 238.90 & $-$15.20 & 28.27 &  &  &  &  &  &  & 2013 \\
 150 & 3 & N & 4.84 & 247.30 & $-$16.90 & 32.85 &  &  &  &  &  &  & 2013 \\
 150 & 4 & N & 6.92 & 251.70 & $-$15.40 & 28.91 &  &  &  &  &  &   & 2013 \\
 150 & 5 & N & 11.00 & 248.20 & $-$10.80 & 21.58 &  &  &  &  &  &  & 2013 \\
\multicolumn{14}{l}{}\\
 150 & 0 & S & $-$1.02 & 258.00 & $-$24.00 & 27.84 &  &  &  &  &  &   & 1994 \\
 150 & 1 & S & $-$3.61 & 257.90 & $-$26.60 & 27.50 &  0.331 &  0.792 &
 120.80 & 244.40 &  4.30 &    7 & 2010 \\
 150 & 8 & S & $-$3.96 & 237.80 & $-$24.20 & 29.00 & 0.383 & 0.849 &
 113.90 & 228.10 &  4.70 & 115 & 2022 \\
\hline
\multicolumn{14}{l}{Daytime $\beta$-Taurids, \#173}\\
 173 & 0 & N & 0.27 &  84.66 &  23.61 & 29.00 &   0.274 &  0.834 &
  52.31 & 101.51 &  0.30 &   41 & 1976 \\
\multicolumn{14}{l}{}\\
 173 & 1 & S & $-$4.40 &  86.74 &  19.01 & 30.00 &  0.340 &  0.850 &
 245.99 & 277.16 &  6.00 &   & 1973 \\
 173 & 2 & S & $-$5.49 &  86.43 &  17.91 & 29.34 &  0.340 &  0.850 &
 223.99 & 278.86 &  6.00 &   & 1954 \\
 173 & 3 & S & $-$3.23 &  82.00 &  20.00 & 27.40 &  0.325 &  0.804 &
 238.30 & 277.00 &  3.60 &  288 & 2008 \\
 \hline
\multicolumn{14}{l}{$\sigma$-Arietids, \#237}\\
 237 & 1 & N & 3.44 &  50.70 &  22.10 & 44.10 &  &  &  &  &  &  475 & 2009 \\
\multicolumn{14}{l}{}\\
 237 & 0 & S & $-$2.74 &  44.69 &  14.10 & 40.48 &  0.110 &  0.980 &
 145.00 &  22.67 &  7.01 &   28 & 1967 \\
\hline
\multicolumn{14}{l}{$h$-Virginids, \#343}\\
 343 & 1 & N & 2.14 & 214.10 & $-$11.40 & 21.35 &  &  &  &  &  &  192 & 2009 \\
\multicolumn{14}{l}{}\\
 343 & 0 & S & $-$1.42 & 204.20 & $-$11.60 & 18.70 &  &  &  &  &  &   16 & 2009 \\
 343 & 2 & S & $-$1.11 & 204.80 & $-$11.50 & 17.20 &  0.742 &  0.659 &
  72.70 & 218.20 &  0.90 &   11 & 2016 \\
 343 & 3 & S & $-$1.35 & 203.60 & $-$11.30 & 18.10 &  0.763 &  0.730 &
  64.10 & 220.40 &  0.60 &  174 & 2020 \\
 343 & 4 & S & $-$1.01 & 203.00 & $-$10.70 & 18.80 &  0.740 &  0.741 &
  66.00 & 219.80 &  0.40 &   34 & 2020 \\
 343 & 5 & S & $-$1.19 & 202.50 & $-$10.70 & 18.90 &  0.745 &  0.749 &
  64.50 & 220.10 &  0.50 &   38 & 2020 \\
 343 & 6 & S & $-$1.17 & 202.80 & $-$10.80 & 18.50 &  0.753 &  0.742 &
  65.50 & 218.70 &  0.60 &  143 & 2020 \\
 343 & 7 & S & $-$1.20 & 203.50 & $-$11.10 & 18.90 &  0.750 &  0.756 &
  65.90 & 219.20 &  0.70 &  61 & 2022 \\
\hline \hline
\end{tabular}
\end{flushleft}
\end{table*}

\section{Discussion on the mixed meteors at a shower identification}
\label{discus}

The incorrect mixture of northern and southern solutions of the same
shower evokes a question whether the meteors with the radiants on both
ecliptical hemispheres are not erroneously mixed when the mean
parameters of an ecliptical shower are calculated. Actually, there can
be found the solutions, in the MDC, with the ecliptical latitude lower
than about $0.5^{\circ}$ and number of meteors equal to several hundreds
or few thousands, so far. If such solutions consisted of meteors with
the radiants in only one ecliptical hemisphere of sky, the size of the
radiant area should be lower than $\sim$$1^{\circ}$. Such a compact
shower is improbable. One can rather suspect that the meteors of both
northern and southern branches were included to the solution.

Of course, in a process of shower identification or various studies of
its dynamics, one does not need to divide the whole streams
into two, northern and southern, branches. Keeping the stream
together can sometimes be useful. For example, we can identify 10
meteors belonging to a particular stream, of which eight particles
belong to the northern branch and two particles to the southern branch.
If we identified the strands separately, the southern branch consisting
of only two particles would scarcely be regarded as a shower. However,
it gains a much larger credibility in the context of whole, ten-particle
stream.

Anyway, when a stream is separated and we want to calculate the mean
parameters, the division to the northern and southern branch is
necessary. No meteor with the radiant in the southern (northern)
ecliptical hemisphere should be a member of northern (southern) shower.

\section{Summary}
\label{summary}

We reminded a reason why two branches of an ecliptic meteor shower are
distinguished in the lists of showers and various studies, although both
branches are caused by meteors of the same - in the interplanetary space
- meteoroid stream.

Furthermore, we pointed out the problem that few showers in the MDC list
are represented by both northern and southern solutions. At the same
time, we pointed out the analogous problem at calculation of mean
parameters of a shower; in the calculation, no meteors with radiant on
the southern (northern) ecliptic hemisphere should be the members of
northern (southern) shower.

In the MDC list, we proposed the corrections to remove the southern
(northern) solutions from the northern (southern) shower. These
corrections will appear in a future version of the MDC list of showers
after they have been discussed and approved by the WG members. 
Specifically, we propose that solution(s):\\
$\bullet$ \#17/6 should be moved to shower \#2 as its another solution;\\
$\bullet$ \#63/0 should be moved to the List of removed showers data;\\
$\bullet$ \#76/0 should be regarded as a new shower and named
``September $\iota$-Aquariids'', with a new IAUNo assigned;\\
$\bullet$ \#96/1 and \#96/5 should become further solutions of
shower \#97;\\
$\bullet$ \#113/0 should be regarded as a new shower and named
``Northern $\alpha$-Cancrids'', with a new IAUNo assigned;\\
$\bullet$ \#113/1 should become another solution of shower \#266,
$\alpha$-Cancrids;\\
$\bullet$ shower $\alpha$-Cancrids, \#266, should be re-named to
``Southern $\alpha$-Cancrids; IAUNo=266 will be retained;\\
$\bullet$ shower Southern $\delta$-Leonids, \#113, should be abolished
(moved to the List of removed showers);\\
$\bullet$ \#115/0,1,5 should be regarded as a new shower and named
``Southern Capricornids-Sagittariids'', with a new IAUNo assigned;\\
$\bullet$ \#115/3,4 should be regarded as a new shower and named
``Northern Capricornids-Sagittariids'', with a new IAUNo assigned;\\
$\bullet$ shower Daytime Capricornids-Sagittariids, \#115, should be
abolished (moved to the List of removed showers);\\
$\bullet$ \#150/2,3,4,5 should be moved to \#149 as its further
solutions;\\
$\bullet$ \#173/0 should be regarded as a new shower and named
``$\beta$-Taurids'', with a new IAUNo assigned;\\
$\bullet$ \#237/0 should become a new shower, which would be named
``Southern $\sigma$-Arietids'', with a new IAUNo assigned;\\
$\bullet$ \#237/1 should become a new shower, which would be named
``Northern $\sigma$-Arietids'', with a new IAUNo assigned;\\
$\bullet$ shower $\sigma$-Arietids, \#237, should be abolished (moved
to the List of removed showers);\\
$\bullet$ \#343/1 should be moved to shower \#21 as its another
solution.\\

The proposed nomenclature changes continue the process of improving the
MDC's base and, as mentioned in the Section \ref{historia}, are in line
with the meteor shower naming rules in force until August 2022,
\citep[see][]{2023NewAR..9601671J}.

\section*{Acknowledgements}

This work was supported, in part, by the VEGA - the Slovak Grant Agency
for Science, grant No. 2/0009/22. This research has made use of NASA's
Astrophysics Data System Bibliographic Services.



\bibliographystyle{elsarticle-harv}
\bibliography{northsouth}{}






\end{document}